\documentclass[twocolumn, twocolappendix]{aastex631}
\usepackage{gensymb}
\usepackage{amsmath}

\allowdisplaybreaks

\accepted{November 24, 2026}

\graphicspath{{./}{figures/}}

\begin{document}

\title{Imprints of the Neutral Interstellar Medium on Polarized Synchrotron Emission and Faraday Rotation}

\author[0000-0002-2679-4609]{Minjie Lei}
\affiliation{Department of Physics, Stanford University, Stanford, CA 94305, USA}
\affiliation{Kavli Institute for Particle Astrophysics \& Cosmology, P.O. Box 2450, Stanford University, Stanford, CA 94305, USA}

\author[0000-0002-7633-3376]{S. E. Clark}
\affiliation{Department of Physics, Stanford University, Stanford, CA 94305, USA}
\affiliation{Kavli Institute for Particle Astrophysics \& Cosmology, P.O. Box 2450, Stanford University, Stanford, CA 94305, USA}

\author[0000-0001-8749-1436]{Mehrnoosh Tahani}
\affiliation{Department of Physics \& Astronomy, University of South Carolina, Columbia, SC 29208, USA}

\author[0000-0003-0932-3140]{A. Bracco}
\affiliation{LUX, Observatoire de Paris, Université PSL, Sorbonne Université, CNRS, 75014 Paris, France}
\affiliation{Laboratoire de Physique de l’École normale supérieure, ENS, Université PSL, CNRS, Sorbonne Université, Université Paris Cité,
Observatoire de Paris, F-75005 Paris, France}
\affiliation{INAF – Osservatorio Astrofisico di Arcetri, Largo E. Fermi 5, 50125 Firenze, Italy}

\author[0000-0003-0742-2006]{Yik Ki Ma}
\affiliation{Max-Planck-Institut für Radioastronomie, Auf dem Hügel 69, 53121 Bonn, Germany}
\affiliation{Research School of Astronomy \& Astrophysics, Australian National University, Canberra, ACT 2611, Australia}

\author[0000-0001-9708-0286]{Amit Seta}
\affiliation{Research School of Astronomy \& Astrophysics, Australian National University, Canberra, ACT 2611, Australia}

\author[0000-0001-7722-8458]{Jennifer West}
\affiliation{Dominion Radio Astrophysical Observatory, Herzberg Research Centre for Astronomy and Astrophysics, National Research Council Canada, PO Box 248, Penticton, BC, V2A 6J9, Canada}
\affiliation{School of Natural Sciences, University of Tasmania, PO Box 807, Sandy Bay, TAS 7006, Australia}

\author[0000-0002-3973-8403]{E. Carretti}
\affiliation{INAF – Istituto di Radioastronomia, via P. Gobetti 101, 40129 Bologna, Italy}



\begin{abstract}

The interstellar medium (ISM) is a complex, multiphase medium, where disentangling the distribution of gas and magnetic field structure across different phases remains a considerable challenge. Recently, Faraday tomography enabled by broadband polarized radio observations has emerged as a promising probe of 3D ISM gas and magnetic field structures. However, the interpretation of these observations is obscured by our limited understanding of the different ISM components probed by the distinct Faraday depth features. In this work, we present a comprehensive multi-frequency ($\sim$300 MHz - 23 GHz) analysis comparing features in the Faraday-rotated, polarized synchrotron emission and \ion{H}{1} structures over the full high-latitude ($|b|>30\degree$) diffuse sky. Using measures of \ion{H}{1} structure complexity along the line of sight (LOS), we observe enhanced depolarization across synchrotron radio frequencies in regions with high \ion{H}{1} complexity characterized by multiple \ion{H}{1} velocity components. We also find that the first and second moments of the Faraday depth spectra are linked to the underlying neutral gas structure. We speculate that these results indicate that the partially ionized neutral medium could directly contribute a significant portion of the diffuse synchrotron emission and Faraday rotation. Photoionized envelopes surrounding neutral clouds and connected magnetic field structure across phases provide possible alternative explanations. These findings establish new observational constraints for Galactic magnetic field models that synthesize multiphase tracers into a single coherent picture.

\end{abstract}


\section{Introduction} \label{sec:intro}

The interstellar medium (ISM) of our Galaxy contains a mixture of gas and dust, threaded by magnetic fields throughout a dynamic and multiphase environment. How gas and magnetic field structures vary across different phases has important implications for a wide range of astrophysical processes that drive stellar and Galactic evolution \citep{pattle23-sc}. However, our current understanding of multiphase ISM structure is limited due to observational constraints. Specifically, different tracers probe gas and magnetic fields in different phases of the ISM, and are limited to partial projections of the true three-dimensional structure. As a result, synthesizing information from different observational tracers into a coherent picture is the key to unlocking a more complete understanding of ISM and Galactic magnetic field structure \citep{beck13-mf,haverkorn15-mf}. 

The ISM can be broadly categorized into ionized and neutral phases across different temperature and column density regimes. For neutral gas phases, the 21 cm hyperfine line is the primary probe of neutral hydrogen \ion{H}{1} structures \citep{hireview23-ah}. In the diffuse ISM at high Galactic latitudes ($|b|>30\degree$), dust grains are well-mixed with \ion{H}{1} gas \citep{boulanger88-ih,lenz17-ln}, with their short axis preferentially aligned with the orientation of the local magnetic fields. As a result, polarized thermal dust emission is an important probe of plane-of-sky (POS) magnetic field structure in the neutral medium \citep{planck18-cr}. Meanwhile, relativistic cosmic ray electrons/positrons spiral around magnetic fields, producing polarized synchrotron emission at radio frequencies. As the polarized synchrotron emission travels through a mixture of (fully or partially) ionized gas and magnetic fields, the polarization angle undergoes Faraday rotation to a degree proportional to the rotation measure (RM) and wavelength squared. RM measurements, derived from radio synchrotron observations at multiple frequency channels, reveal information about the LOS magnetic field \citep{haverkorn19-ft}. 

Correlation studies of different observational tracers has been shown to be an effective way to constrain the variation of multi-phase ISM and magnetic field structures \citep{kalberla16-hm, hensley22-cr, campbell22-cm, lei24-do, halal24-lb}. With advancements in both broadband surveys of the low frequency radio sky \citep{vaneck18-pw}, and the understanding of how magnetic fields influence the morphology of cold \ion{H}{1} \citep{clark18-ch, clark19-nh, peek19-cl}, an active area of inquiry is the degree to which the observed Faraday features at radio frequencies correlate with the neutral tracers, or whether they probe different phases with distinct magnetic field structures. The answers will shed light on the physical nature of the Faraday features and the multiphase structure of Galactic magnetic fields. Results from the LOFAR survey \citep{zaroubi15-ni, kalberla17-ni, jelic18-ni, bracco20-ni, boulanger24-wm, ercey24-ft} found intriguing correlations between the orientation of LOFAR Faraday-rotated diffuse synchrotron emission features, the magnetic field traced by polarized dust emission, and HI filaments over specific regions of the sky. However, this type of direct morphological correspondence seems largely limited to LOFAR frequencies, confined to a few square degree-sized fields or limited regions of the northern sky, and not ubiquitously observed. 

In this study, we conduct a comprehensive multi-frequency ($\sim$300 MHz -- 23 GHz) comparison between polarized synchrotron+Faraday rotation measurements and \ion{H}{1} structures over the full high-latitude ($b|>30\degree$) diffuse sky. We utilize both narrow-band radio polarization datasets, as well as broad-band observations from the Global Magneto-Ionic Medium Surveys \citep[GMIMS,][]{wolleben19-gs, wollenben21-gn, sun25-gm} that enable rotation measure (RM) synthesis analysis of polarized radio observations \citep{Burn66-sb, brentjens05-fr}.  By leveraging RM synthesis, which allows for the decomposition of polarized synchrotron emission components along a line of sight (LOS) into their Faraday depth ($\phi$) contributions, we can examine the potential connections between the LOS information encoded in both the Faraday spectra and the \ion{H}{1} spectra over wide regions across the high-latitude sky. 

The rest of the paper is organized as follows: In Section \ref{sec:data}, we describe the observational datasets used in this work, including the derivation of \ion{H}{1} complexity maps, and various narrow- and wide-band synchrotron datasets. Sections \ref{sec:dust_hi} and \ref{sec:fara_hi} examine the similar imprints of \ion{H}{1} complexity on both polarized dust emission and synchrotron emission maps. In Sections \ref{sec:connect_m1} - \ref{sec:connect_m2}, we present our results regarding the connection between neutral gas LOS complexity and Faraday rotation features, focusing on rotation measure ratios and Faraday spectrum moments. We discuss the physical interpretation of these correlations in Section \ref{sec:discussion}, and provide our concluding remarks in Section \ref{sec:conclusion}. 

\begin{figure*}[t]
    \centering
    \includegraphics[width=0.95\textwidth]{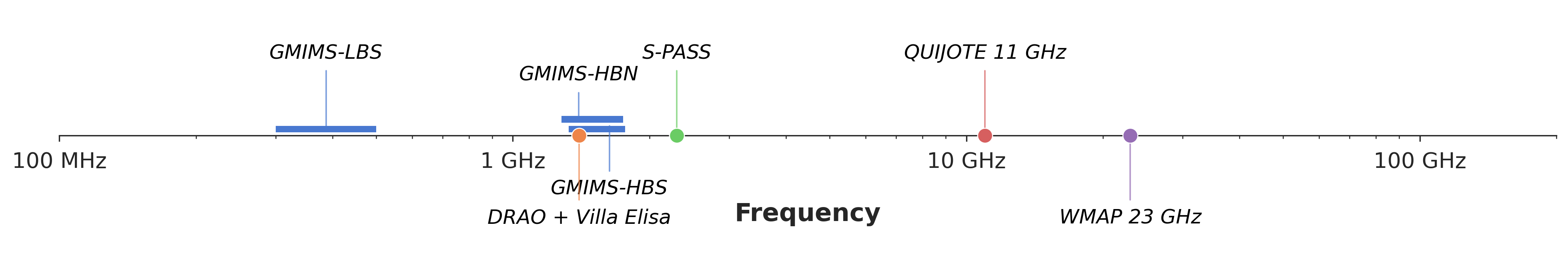}
    \caption{Frequency coverage of the polarized synchrotron emission surveys analyzed in this paper.}
    \label{fig:survey_freqs}
\end{figure*}

\section{Data} \label{sec:data} 

\subsection{Map of \ion{H}{1} Complexity} \label{subsec:hi_nc}
To measure the complexity of \ion{H}{1} structures along the LOS, we employ the map developed by \citet{panopoulou20-nc}. The authors applied Gaussian decomposition to identify distinct emission components in the \ion{H}{1} emission spectra from the \ion{H}{1}4PI survey \citep{hi4pi16-rl}. Only common components identified by multiple neighboring pixels are retained to ensure spatial coherence. The decomposition is applied to the high latitude sky where the neutral hydrogen is mostly in the atomic phase, with a column density mask of $N_\mathrm{H\,I}<4\times10^{20}~\mathrm{cm^{-2}}$. The result is a map of the number of \ion{H}{1} clouds along the LOS in the high latitude diffuse sky. To account for column density variation between clouds, \citet{panopoulou20-nc} additionally defined a LOS complexity measure weighted by the column density of the components
\begin{equation} \label{eq:nc}
    \mathcal{N}_c = \sum_{i=1}^{N_\mathrm{clouds}}\frac{N_\mathrm{H\,I}^i}{N_\mathrm{H\,I}^\mathrm{max},}
\end{equation}
where $N_\mathrm{H\,I}^i$ is the column density of the $i$th cloud along the sightline, and $N_\mathrm{H\,I}^\mathrm{max}$ is the maximum cloud column density for a given sightline. In the case where the total column density of the sightline is dominated by a single component, $\mathcal{N}_c$ will be $\sim 1$. On the other hand, a sightline with $n$ equal-column-density clouds will have $\mathcal{N}_c=n$. Higher $\mathcal{N}_c$ is primarily driven by the presence of multiple intermediate velocity clouds along the LOS \citep{panopoulou20-nc}, and displays a trend of higher complexity in the Northern hemisphere than the Southern hemisphere. 

\subsection{Polarized Dust Emission} \label{subsec:pol_dust}

For polarized dust emission, we use the 80\arcmin\ Planck PR3 data at 353 GHz. Following the fiducial zero-level correction adopted by \citet{planck18-cr}, we subtract a CIB monopole correction of 452 $\mu\mathrm{K_{CMB}}$, then add a Galactic offset correction of 63 $\mu\mathrm{K_{CMB}}$. From the offset-corrected Stokes $I, Q, U$ maps and their covariances, we apply the modified asymptotic estimator method introduced in \citet{plaszczynski14-db} to arrive at the noise-debiased polarized intensity map $P_{353}$ and its uncertainty estimate $\sigma_{P_{353}}$. We retain the fraction of the sky where the signal-to-noise ration (SNR) is $P_{353}/\sigma_{P_{353}}>3$, which amounts to 96\% of the original \ion{H}{1} complexity $N_c$ map footprint. Finally, to match the resolution at 80\arcmin, we downgrade the Planck maps from their native HEALPix pixelization at $N_{\rm side}=2048$ to $N_{\rm side}=64$.

\subsection{Wide-band Polarized Synchrotron Emission} \label{subsec:pol_wide}
We make use of publicly available polarized radio emission datasets from the Global Magneto-Ionic Medium Survey \citep[GMIMS;][]{gmims09-fd} that map the full sky in wide frequency bands from $\sim300$ MHz to $\sim1700$ MHz, enabling the characterization of Faraday rotation. Here we summarize each GMIMS dataset that we use (see Figure \ref{fig:survey_freqs}).

\begin{itemize}
    \item \textbf{GMIMS High Band North Survey (GHBN)} maps the northern sky with the Dominion Radio Astrophysical Observatory (DRAO) 26m telescope from declinations $-30\degree$ to $+87\degree$ at all right ascensions with angular resolution 40\arcmin\ \citep{wollenben21-gn}. GHBN covers the frequency range 1280 to 1750 MHz, with a channel width of 0.24 MHz.
    \item \textbf{GMIMS High Band South survey (GHBS)} covers the whole sky south of the declination of $0\degree$ over frequency range $1328-1768$ MHz with a channel width of 1 MHz with Murriyang, the Parkes 64-m telescope \citep{sun25-gm}.
    \item \textbf{GMIMS Low Band South survey (GLBS)} maps the sky between declinations $-90\degree$ and $20\degree$. The GLBS utilizes the CSIRO Parkes 64-m Telescope to survey frequencies between 300 to 480 MHz, with angular resolution varying between $80\arcmin$ and $45\arcmin$, and 0.5 MHz channel width \citep{wolleben19-gs}.
\end{itemize}

We smooth the Stokes Q and U maps of each of the GMIMS datasets to $80\arcmin$ and and apply RM-CLEAN \citep{Heald2009} and RM synthesis \citep{brentjens05-fr} to derive Faraday depth cubes following the same procedures in \citet{wollenben21-gn}, using the RM-Tools package from the Canadian Initiative for Radio Astronomy Data Analysis \citep[CIRADA;][]{pucell05-rm, vaneck26-rt}. For the GHBN and GHBS datasets, this results in a Faraday depth spectrum $P(\phi)$ covering the Faraday depth range $-500  <\phi<500 \mathrm{\;rad\;m^{-2}}$ in bins of 5 $\mathrm{rad\;m^{-2}}$. For the low-frequency GLBS, $P(\phi)$ spans $-100 <\phi<100 \mathrm{\;rad\;m^{-2}}$ in steps of 0.5 $\mathrm{rad\;m^{-2}}$. 

To estimate the uncertainty level of the Faraday depth spectrum, we follow the procedure outlined in \citet{raycheva25-gm}. A noise map $\sigma$ is derived by fitting a Rayleigh distribution $R(x;\sigma)=\frac{x}{\sigma^2}e^{-x^2/2\sigma^2}$ to the Faraday depth spectrum $P(\phi)$ in the range $|\phi|\in[750,1000]\mathrm{\;rad\;m^{-2}}$ for GHBN and GHBS, and $|\phi|\in[100,200]\mathrm{\;rad\;m^{-2}}$ for GLBS where we expect instrumental noise to dominate. 

Finally, we reduce the Faraday depth cube data by computing the moment maps of $P(\phi)$, following the definition in \citet{dickey19-mm}:
\begin{align}
    M_0 &= \Delta\phi \sum_{i=1}^{n} P_i (\text{K rad m}^{-2} ) \label{eq:M0} \\ 
    M_1 &= \frac{\Delta\phi \sum_{i=1}^{n} P_i \phi_i}{M_0} (\text{rad m}^{-2}) \label{eq:M1} \\  
    M_2 &= \frac{\Delta\phi \sum_{i=1}^{n} P_i (\phi_i - M_1)^2}{M_0} (\text{rad m}^{-2})^2 \label{eq:M2}
\end{align}
where $n$ is the number of Faraday depth channels. $\phi_i$ and $P_i$ are the Faraday depth and polarized intensity at channel $i$ respectively. $\Delta \phi$ is the width between Faraday depth channels, which is 5 $\mathrm{\;rad\;m^{-2}}$ for GHBN and 0.5 $\mathrm{\;rad\;m^{-2}}$ for GLBS. Intuitively, the 0th moment gives the total polarized intensity integrated over Faraday depth; $M_1$ is the intensity-weighted mean Faraday depth; and $M_2$ measures the dispersion of the Faraday depth spectrum from the weighted mean $M_1$. We make use of the existing GHBS Faraday moment maps computed by \citet{raycheva25-gm}, and utilize the same procedure to produce maps for GHBN and GLBS. 

\subsection{Narrow-Band Polarized Synchrotron Emission} \label{subsec:pol_narrow}
To complement our analysis with broadband datasets over frequencies $\sim0.3$ to 1 GHz from the GMIMS surveys, we make use of several narrow-band polarized radio emission data at higher frequency ranges from $\sim1$ to 20 GHz. 

\begin{itemize}
    \item \textbf{The Wilkinson Microwave Anisotropy Probe (WMAP)} survey maps all-sky total intensity and polarization in five narrow frequency bands from 23 GHz to 94 GHz. We make use of the lowest-frequency $K$-band data, centered at 23 GHz. To quantify instrumental uncertainty, we follow the description in \citet{wmap13-dn} to compute the pixel noise in $mK$:
    \begin{align}
        \sigma=\sigma_0/\sqrt{N_{\rm obs}}
    \end{align} 
    where $\sigma_0$ is the noise root mean square (rms) for one observation, and $N_{\rm obs}$ is the number of observations taken at each pixel. For WMAP K-band data, $\sigma_0=1.429$ for Stokes $I$ and $\sigma_0=1.435$ for Stokes $Q$ and $U$ maps. We use the $N_{\rm obs}$ values provided in the full resolution coadded nine year K band maps, which ranges from $\sim400$ to $\sim20000$. Using the resulting $\sigma$ map as the standard derivation, we generate Gaussian noise samples over 1000 trials to propagate the uncertainty to the polarized intensity map. 
    \item \textbf{Q-U-I JOint Tenerife Experiment (QUIJOTE)} surveys the northern sky in four frequency bands from 11 GHz to 19 GHz with angular resolution of $\sim 1\degree$ \citep{quijote23-sv}. We utilize the lowest-frequency IQU maps, centered at 11 GHz. We characterize the noise level of the dataset from the difference of the two half-mission maps:
    \begin{align}
        \sigma=\frac{h_{1}-h_{2}}{w}
    \end{align}
    where $h_1$ and $h_2$ are the two half-mission maps, and the normalizing weight $w$ is computed as:
    \begin{align}
        w=\sqrt{\left(w_{1}+w_{2}\right)\left(w_{1}^{-1}+w_{2}^{-1}\right)}
    \end{align}
    $w_1$ and $w_2$ are individual weight maps of the half-mission maps respectively and provided as part of the half-mission datasets. We again use the $\sigma$ map as the standard derivation, and generate Gaussian noise samples over 1000 trials to propagate the uncertainty level to the final polarized intensity map. 
    \item \textbf{S-Band Polarization All Sky Survey (S-PASS)} is a polarized radio emission survey of the southern sky over DEC $< -1\degree$ at a frequency of 2.3 GHz \citep{spass19-sv}. The survey is carried out with the Parkes radio telescope at an angular  resolution of 8.9 arcmin. We make use of the Stokes $IQU$ as well as the pixel sensitivity map $\sigma_{\rm px}$. 
We estimate the overall uncertainty level by adding the statistical uncertainty ($\sigma_{\rm pix}$) with a map level calibration uncertainty of $5\%$ in quadrature \citep{spass19-sv}. Using the resulting uncertainty map as the standard deviation, we generate Gaussian noise realizations over 1000 trials to estimate uncertainty level of the final polarized intensity map.
    \item \textbf{The DRAO+Villa Elisa polarization maps \footnote{\url{https://cade.irap.omp.eu/dokuwiki/doku.php?id=drao}}} cover the full sky at 1.4 GHz by combining the northern sky survey carried out by the 25.6 m DRAO telescope in Canada \citep{drao06-ns} with the southern sky survey by the 30m Villa Elisa telescope in Argentina \citep{villaelisa08-ss}, both at $36\arcmin$ angular resolution. We adopt the map-level sensitivity of 15 mK as the standard deviation. 
\end{itemize}

From all the narrow-band polarization datasets, we compute the debiased polarized intensity from the Stokes IQU and uncertainty map $\sigma$ as:
\begin{align}
    P=\sqrt{Q^{2}+U^{2}-\sigma^{2}}
\end{align}
Before computing the $P$ map, we smooth the stokes $Q$ and $U$ maps to 80\arcmin\ resolution and downgrade the maps to $N_{\rm side}=64$.
\subsection{Rotation Measures of Extragalactic Sources} \label{subsec:rm_ex}
While the GMIMS surveys map the Faraday rotation measure of diffuse Galactic synchrotron emission, another method to survey the Galactic rotation measure sky is to map the polarized emission of compact extragalactic sources. Here we utilize the Galactic Faraday Rotation sky map from \citet{hutsch22-fm}. The RM sky map is created from compiling, gridding, and interpolating $\sim50000$ data sources from 41 surveys using a Bayesian inference scheme. From the extragalactic RM measurements, which contain contribution from the Galactic foreground, the intrinsic Faraday rotation of the source, and everything in between, \citet{hutsch22-fm} utilize the Bayesian inference scheme to separate and estimate the Galactic foreground RM contribution as a smooth function. We use the Galactic Faraday Rotation sky map and compare it to the GMIMS $M_1$ maps in overlapping regions. 

\begin{figure*}[t]
    \centering
    \includegraphics[width=0.95\textwidth]{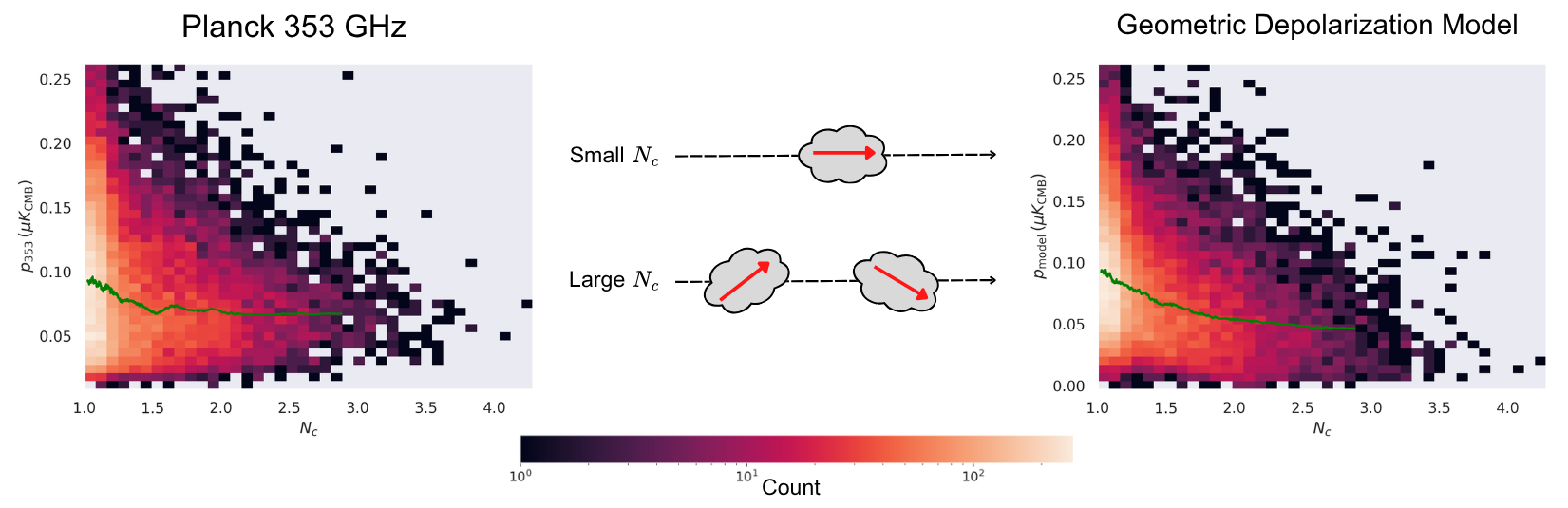}
    \caption{Comparison of the polarization fraction of dust emission from Planck data (left) and our cartoon model (right) vs. \ion{H}{1} complexity from \citet{panopoulou20-nc}. The green line indicates the running median in equal sightline bins for each distribution. The cartoon model qualitatively reproduces the observed trend of decreasing polarized fraction with increasing \ion{H}{1} complexity.}
    \label{fig:cartoon_model_dust}
\end{figure*}

\section{Imprint of \ion{H}{1} complexity on Polarized Dust Emission} \label{sec:dust_hi}
Recent studies have shown strong evidence that variation in polarized dust emission at large angular scales over the diffuse sky is driven by the 3D geometry of the Galactic magnetic field. \citet{planck18-cr} found that the polarization fraction at 353 GHz is anti-correlated with the local polarization angle dispersion at 160\arcmin\ resolution. They further showed that this relationship is consistent with models of turbulent magnetic field structure that assumes uniform intrinsic dust alignment properties. This indicates that magnetic field structure alone is able to reproduce the main statistical variation of the observed dust polarization. Utilizing 3D mapping of dust extinction information in the solar neighborhood up to a distance of $\sim1.25$ kpc \citep{edenhofer24-dm}, \citet{halal24-lb} measured the number of coherent dust components along each sightline, and showed that the dust polarization fraction at 353 GHz decreases towards more complex sightlines with multiple dust components, compared to simple sightlines at equivalent total extinction. This suggests that polarized dust emission is sensitive to variations in the magnetic field geometry along the LOS sampled by the extended dust distribution.

In the high latitude ($|b|>30\degree$), diffuse ISM, the \ion{H}{1} column density is linearly correlated with dust emission \citep{boulanger88-ih, lenz17-ln}. \ion{H}{1} and dust emission can thus be used as complementary probes of the diffuse neutral medium. 
\citet{clark18-ch} analyzed the coherence of the orientation of \ion{H}{1} structures as a function of their line-of-sight velocity, and found that \ion{H}{1} coherence is positively correlated with the polarization fraction of the Planck 353 GHz dust emission.  As described in Section \ref{subsec:hi_nc}, \citet{panopoulou20-nc} developed a metric to quantify the complexity of \ion{H}{1} velocity structures along the LOS, and showed that the dust polarization fraction decreases towards more complex sightlines with multiple \ion{H}{1} components, in diffuse regions where $N_\mathrm{H\,I}<4\times10^{20}~\mathrm{cm^{-2}}$. This imprint of LOS \ion{H}{1} complexity on polarized dust emission is consistent with geometric depolarization, where sightlines with multiple \ion{H}{1} components are more likely to sample differently oriented magnetic fields along the LOS, leading to additional depolarization compared to sightlines dominated by a single \ion{H}{1} component. 

We illustrate this argument with a simple cartoon dust polarization model. Dust polarization fraction $p$ can be written as: 
\begin{equation} \label{eq:p_model}
    p_{\rm model} =\bigg|\int p_0(s) \cdot{e^{2i\psi(s)}ds}\bigg|
\end{equation}
where $\psi$ is the plane-of-sky polarization angle. $p_0$ is the intrinsic polarization fraction, which we take to represent all other factors contributing to the $p$ variation, such as dust emissivity, grain alignment, and polarizing efficiency. We adopt a data-driven approach to estimate $p_0$ by fitting a log-normal distribution to Planck 353 GHz polarized fraction data in regions where $N_c<1.01$, i.e. sightlines compatible with a single \ion{H}{1} component. We assume that the $N_c<1.01$ sightlines samples the distribution of intrinsic $p_0$ variation without the effect of LOS complexity. Then for each sightline, we sample a $p_0$ from the fitted distribution, and generate a random polarization angle $\psi$ from a uniform distribution $U(0,\pi)$ for each of the cloud components identified by the \citet{panopoulou20-nc} method, weighted by the column density of the cloud. The resulting model polarization fraction $p_{\rm model}$ is shown in Figure \ref{fig:cartoon_model_dust}, in comparison to the observed Planck 353 GHz polarized fraction data $p_{353}$. Figure \ref{fig:cartoon_model_dust} shows that this simple geometric depolarization model is able to qualitatively reproduce the observed $p$-$N_c$ trend, showing a decreasing dust polarization fraction with increasing \ion{H}{1} complexity. 

\begin{figure*}[t]
    \centering
    \includegraphics[width=0.9\textwidth]{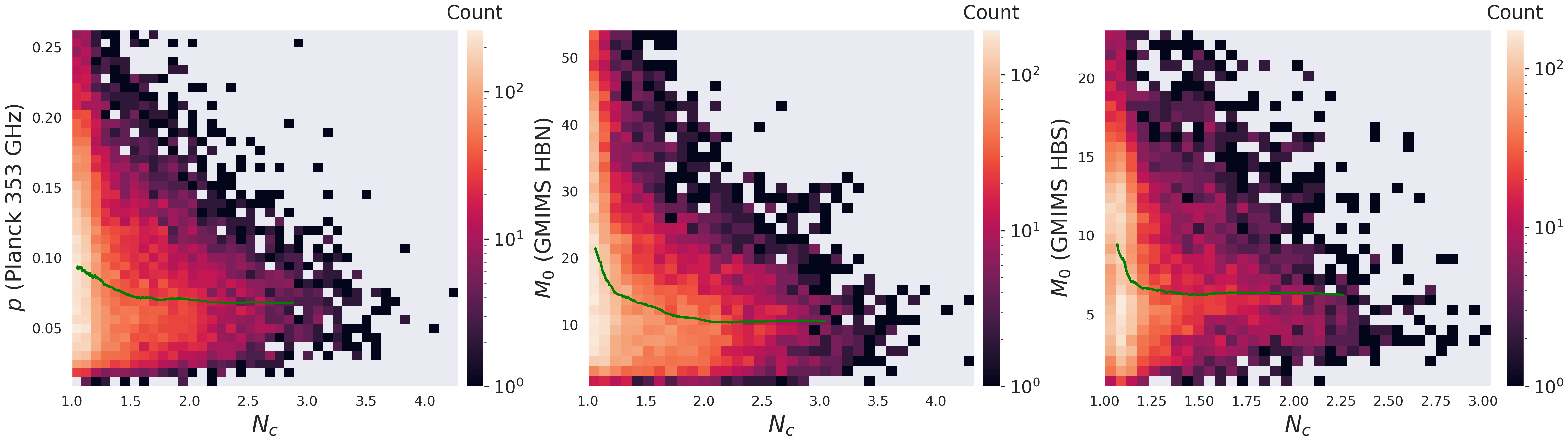}
    \caption{2D histogram of polarization emission from GHBN and GLBS surveys vs. \ion{H}{1} complexity ($N_c$), compared to the dust $p_{353}-N_c$ relationship. Similar to dust emission, polarized synchrotron data shows decreasing polarized emission with increasing \ion{H}{1} complexity for both the northern and southern sky footprints.}
    \label{fig:p_nc_gmims}
\end{figure*}

\section{Imprint of \ion{H}{1} complexity on Polarized Synchrotron Emission} \label{sec:fara_hi}

We now examine the relationship between \ion{H}{1} complexity and polarized synchrotron emission. In Figure \ref{fig:p_nc_gmims}, we compare the 0th moment $M_0$ from the GHBN and HBS surveys vs. \ion{H}{1} complexity. Both the HBN and HBS surveys cover frequencies between $\sim1300-\sim1700$ MHz, over the northern and southern sky respectively. As described in Section \ref{subsec:pol_wide}, the 0th moment $M_0$ is total polarized intensity integrated over Faraday depth, or equivalently, over the frequency range covered by the surveys. Similar to the dust polarization result in Figure \ref{fig:cartoon_model_dust} and in \citet{panopoulou20-nc}, the 0th moment polarized intensity shows a decreasing trend with increasing \ion{H}{1} complexity for both the northern and southern sky footprints. The $M_0$-$N_c$ relationship is stronger for the northern sky HBN survey. This can be attributed to the higher dynamic range of $N_c$ in the northern sky. The 90th percentile $N_c$ value is 2.25 for the northern footprint, and only 1.65 for the southern footprint. The higher complexity over the HBN footprint likely reflects the larger population of intermediate velocity clouds (IVCs) found over the northern sky \citep{danly89-iv, welsh04-iv, panopoulou20-nc}. 

The analogous relationship between polarized synchrotron and dust emission vs. $N_c$ raises the intriguing possibility that the LOS complexity of \ion{H}{1} clouds also traces the magnetic complexity of synchrotron emitting structures. This interpretation would be consistent with existing results showing a direct morphological correlation between \ion{H}{1} and synchrotron structures in specific regions \citep{zaroubi15-ni, kalberla17-ni, vaneck17-ph,jelic18-ni, bracco20-ni, boulanger24-wm}, and suggests that there may be a broader statistical correlation between these tracers. 
However, there are a few caveats to the comparison between the polarized dust emission in Figure \ref{fig:cartoon_model_dust} and the polarized synchrotron emission in Figure \ref{fig:p_nc_gmims}. First, $M_0$ measures the total polarized intensity $P$ and not polarization fraction $p=P/I$. The polarization fraction is a better quantity for this comparison, because we wish to isolate and constrain the effect of possible geometric depolarization, and remove the effect of total intensity variation. For radio surveys like GMIMS, the zero level of the total intensity $I$ is affected by different radio contamination and carries substantial systematic error. The GHBN and HBS surveys in particular do not attempt to correct for zero level offsets in total intensity. As a result, rather than analyzing the polarization fraction directly, we devise a statistical test that avoids this issue (Section \ref{subsubsec:depol_measure}). 

Another caveat to comparing the polarized dust and synchrotron emission results is that for polarized synchrotron emission data, in addition to geometric depolarization, there can be significant, frequency-dependent depth depolarization due to Faraday rotation. Taking into account Faraday rotation, the polarization fraction formula in Equation \ref{eq:p_model} generalizes to:
\begin{equation} \label{eq:p_general}
  p(\lambda) =
  \frac{\left|\displaystyle\int \epsilon(s,\lambda)\,p_0(s,\lambda)
  e^{2i[\psi_0(s)+\phi(s)\lambda^2]}\,ds\right|}
  {\displaystyle\int \epsilon(s,\lambda)\,ds}
\end{equation}
where $\epsilon$ is the synchrotron emissivity, and the rotation measure $\phi(s)$ is given by:
\begin{equation} \label{eq:phi}
    \phi(s)=0.81\int^s_0\frac{n_e(s')}{\mathrm{cm^{-3}}}\frac{B_{LOS}(s')\cdot ds'}{\mathrm{\mu G\,pc}}\,\mathrm{rad\,m^{-2}}
\end{equation}
where $n_e$ is the electron density, and $B_{LOS}$ is the LOS magnetic field component. As a result of the $\lambda^2$ dependency, depth depolarization becomes more significant for lower-frequency synchrotron surveys. A fully consistent explanation for any observed relationship between polarized synchrotron emission and \ion{H}{1} complexity would require a careful comparison of datasets at different frequencies, which we address in the next section. 

\begin{figure*}[t]
    \centering
    \includegraphics[width=0.95\textwidth]{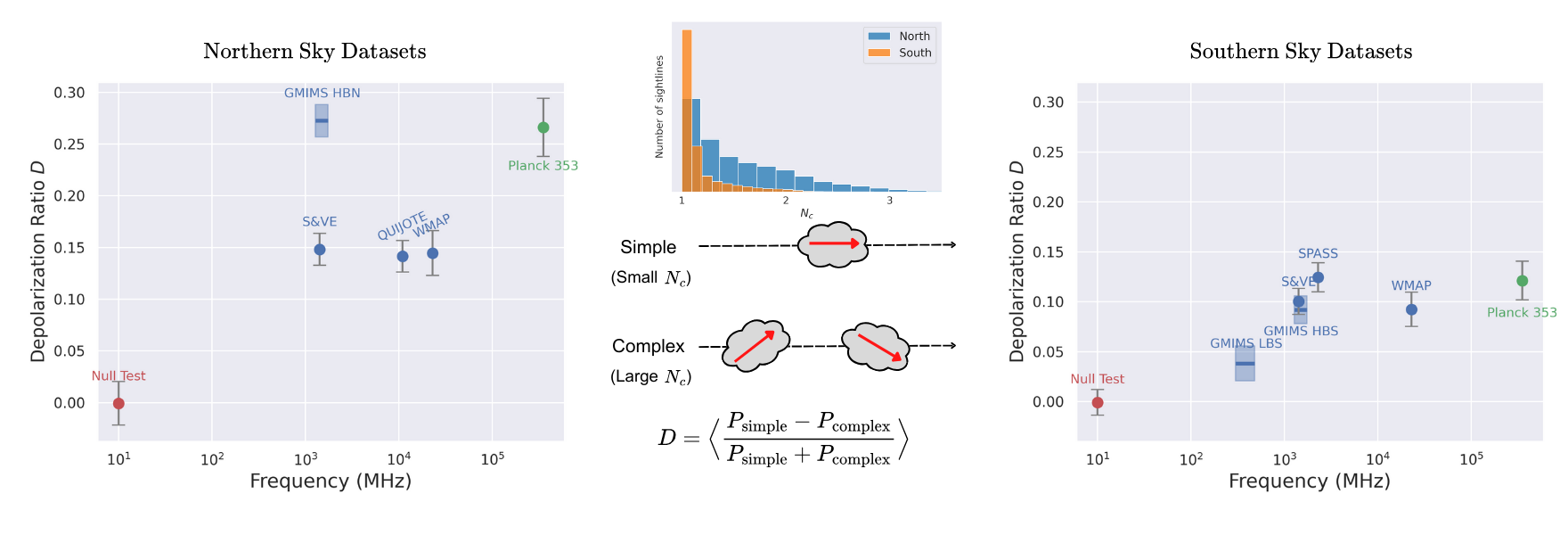}
    \caption{Relative depolarization between \ion{H}{1} simple and complex sightlines as measured by the depolarization ratio $D$ defined in Equation \ref{eq:depol_metric}, for synchrotron datasets across multiple frequencies, compared to Planck dust emission at 353 GHz, over northern (left panel) and southern (right panel) footprints. Total synchrotron intensity $I$ is treated as the confounding variable when pairing low and high complexity sightline. The depolarization ratios for contrasting LOS complexity sightlines are compared to null tests where $D$ is computed over two randomly drawn sightline groups over the sky.}
    \label{fig:depol_full}
\end{figure*}

\subsection{Methodology} \label{subsec:method}

\subsubsection{Depolarization Ratio} \label{subsubsec:depol_measure}

To quantify the relationship between LOS HI complexity and synchrotron polarization fraction without directly dividing polarized intensity data by total intensity with uncertain zero levels, here we define a depolarization ratio $D$ using a nearest-neighbor matching process introduced in \citet{halal24-lb}. Nearest neighbor matching is commonly used in causal inference problems to estimate causal relationships while controlling for confounding variables \citep{Stuart10-nn}. We divide the sky into two groups, one with low \ion{H}{1} complexity ($\leq$10th percentile $N_c$), and the other with high \ion{H}{1} complexity ($\geq$90th percentile $N_c$). We experimented with various different $N_c$ limits including (20th percentile, 80th percentile) and obtained qualitatively similar results. We treat the total synchrotron emission intensity $I$ as the confounding variable when comparing the distribution of $M_0$ in the low complexity vs. high complexity groups: we match pairs of sightlines, one from each complexity group, by the closest total $I$. From each matched sightline pair, we subtract $M_0$ of the high complexity group from that of the low complexity group, and normalize over their sum to arrive at the depolarization ratio $D$:
\begin{equation} \label{eq:depol_metric}
    D=\underset{\rm pairs}{\mathrm{Ave}}\biggl\{\frac{M_0^{\rm low}-M_0^{\rm high}}{M_0^{\rm low}+M_0^{\rm high}}\Biggr\}
\end{equation}
Defined this way, a positive $D$ value indicates that the sightlines with high \ion{H}{1} complexity are more depolarized on average than sightlines with low \ion{H}{1} complexity. The normalization over the sum of the low vs. high complexity pair rescales the metric to be between -1 and 1, and enables comparison between datasets at different frequencies with different total and polarized intensity ranges. 

\subsubsection{Statistical Significance Test} \label{subsubsec:stat_test}
To determine the statistical significance of our LOS complexity depolarization results and to compare datasets across multiple frequencies, we utilize a set of permutation tests. For each dataset $i$ described in Section \ref{sec:data}, we perform the analysis outlined in Section \ref{subsubsec:depol_measure} to compute a depolarization measure $D_i$. To estimate the uncertainty $\Delta D_i$, we rerun the analysis using bootstrap sampling with replacement for 500 runs, and compute the mean and standard deviation from the samples. 

We also perform a null test $D_{\rm null}$ to validate the statistical significance of our depolarization measure results. Instead of dividing the sky into two groups of high vs. low complexity, for the null test we randomly select 20\% of the sky to be in group 1, and another randomly selected 20\% in group 2. These random selections are then pair-matched according to total $I$. The uncertainty $\Delta D_{\rm null}$ is estimated using the same bootstrap sampling procedure. The expectation $\left<D_{\rm null}\right>$ should approach zero for unbiased random selection. When comparing depolarization tests to the null test, we treat results with a two-tailed p-value $< 0.001$ as statistically robust.

\subsection{Result: Evidence for LOS Complexity Depolarization Across Multi-Frequency Datasets} \label{subsec:fara_hi_freq}

Figure \ref{fig:depol_full} presents the depolarization ratio comparison for each of our datasets and for the null test, treating the northern and southern skies separately. In the left panel of Figure \ref{fig:depol_full}, we show the result of the depolarization ratio analysis for multi-frequency datasets in the northern sky footprint (Planck 353 GHz,  WMAP, QUIJOTE, DRAO+Villa Elisa, GHBN), compared to the null test result $D_{\rm null}$. The full-sky surveys are masked to match the overlapped footprint of the northern sky surveys GHBN and QUIJOTE. The 10th and 90th percentile limits of \ion{H}{1} complexity $N_c$ used to select simple and complex sightlines are 1.05 and 2.24 respectively, approximately representing a sightline dominated by a single cloud versus a sightline dominated by two equal-column-density clouds. 

The null test result is consistent with zero, with $D_{\rm null}=-0.004\pm0.021$. The two-tailed p-value for all datasets compared to the null test is less than 0.001 and statistically significant when compared to the null test. There is a consistent relative depolarization for \ion{H}{1} complex sightlines at the level of  $D\sim0.15$ across synchrotron frequencies from 1.4 GHz to 23 GHz. The exception is GMIMS-HBN (1.28-1.75 GHz), which has a higher level of relative depolarization at $D\sim0.25$, comparable to that of the polarized dust emission data at 353 GHz. This is likely related to the unexpectedly high level of polarized intensity observed in GMIMS-HBN. \citet{raycheva25-gm} reported that the level of PI in GMIMS-HBN is almost twice that of GMIMS-HBS in overlapping regions. 

A caveat for the results at the frequencies of the QUIJOTE and WMAP surveys is that anomalous microwave emission (AME) can account for significant fractions of the total intensity \citep{mdma08-gm}. If the AME contribution to total intensity differs significantly in regions of low $N_c$ versus regions of high $N_c$, then it could bias the sightline matching based on total intensity even though the AME contribution to polarized emission is expected to be minimal \citep{herman-am}. To investigate this effect, we follow the procedure outlined in \citet{hensley22-cr}, which examined and found no relationship between AME emission and CNM fraction, using the AME maps derived from the \textsc{Commander} component separation of the combined Planck and WMAP observations of the microwave sky \citep{planck16-sp}. Utilizing the posterior mean value of the \textsc{Commander} fits at $1\degree$ resolution and $N_{\rm side}=256$, we construct maps of the total AME intensity at 11 GHz and 23 GHz, corresponding to QUIJOTE and WMAP specifically, smoothed to 80\arcmin\ and downgraded to $N_{\rm side}=64$. We found no correlation between the ratio of AME to total intensity and $N_c$. Furthermore, the depolarization ratio results are consistent when including AME intensity maps as an additional confounding variable, or if we repeat our analysis after subtracting the AME maps from the QUIJOTE and WMAP total intensity maps.

The lower level of relative depolarization at synchrotron frequencies compared to dust at 353 GHz can be attributed to several factors. First, we do not expect that the synchrotron-emitting cosmic rays are as tightly correlated with the \ion{H}{1} as the dust is: the relativistic electrons are likely much more volume-filling than the neutral medium. 
Moreover, Faraday rotation effects start to play an increasingly significant role at lower frequencies, due to the frequency-dependent depth depolarization described by Equation \ref{eq:p_general}. At low frequencies and high $\lambda^2$, a significant portion of the sightline becomes depolarized. \citet{uyaniker03-ph} described this effect as a ``polarization horizon", a $\lambda^2$-dependent distance beyond which all emission is depolarized. More recent work \citep{hill18-ph} has argued that the depth depolarization effect does not introduce a hard horizon. Nevertheless, at low frequencies the polarized synchrotron emission may not trace as much of the sightline probed by the \ion{H}{1} complexity as the high frequency synchrotron data due to greater depth depolarization. 

\begin{figure}[!htbp]
    \centering
    \includegraphics[width=\columnwidth]{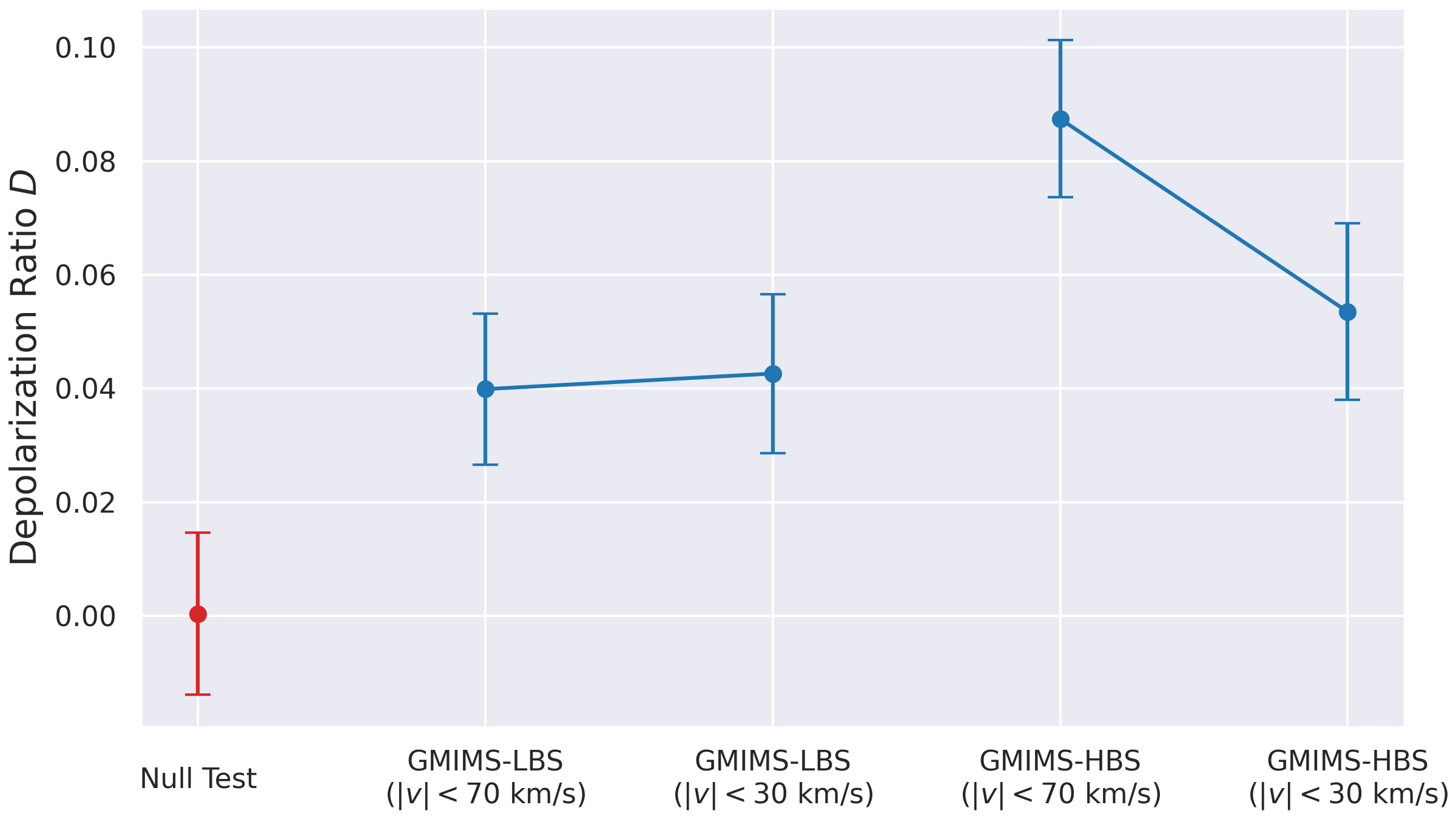}
    \caption{Comparison between the relative depolarization ratio of the GMIMS-LBS ($300-480$ MHz) and GMIMS-HBS ($1328-1768$ MHz) datasets, when defining LOS complexity using different \ion{H}{1} cloud velocity bounds. While relative depolarization decreases significantly for high frequency data when \ion{H}{1} complexity is confined to a smaller velocity range and thus a smaller portion of the sightline, it remains consistent for the low frequency synchrotron data. }
    \label{fig:depol_nc_vrange}
\end{figure}

This frequency-dependent effect is better reflected in the right panels of Figure \ref{fig:depol_full} for datasets in the Southern sky footprint (Planck 353 GHz, WMAP, SPASS, DRAO+Villa Elisa, GMIMS-HBS, GMIMS-LBS). The overall scale of the depolarization ratio $D$ over the southern sky is lower than that over the northern sky footprints, due to the smaller dynamic range of $N_c$ in the southern footprint. The 10th and 90th percentile limits of $N_c$ are 1.05 and 1.65, respectively. In other words, the ``complex" sightlines in the Southern hemisphere are not generally as complex as their counterparts in the Northern hemisphere. The depolarization ratio of GMIMS-HBS is at $D\sim0.10$, consistent with other similar frequency synchrotron datasets in the southern hemisphere. In comparison, the low-frequency GMIMS-LBS has a much lower $D\sim0.04$ but still statistically significant (two-tailed p-value $< 0.001$). 

To further test the hypothesis that the frequency dependence of D in the radio data is driven by the frequency dependence of depth depolarization, we conduct an experiment where the depolarization ratio is computed with an alternative version of the $N_c$ metric that probes a smaller extent of the LOS. Instead of the full $|v|< 70$ km/s range limit described in Section \ref{subsec:hi_nc} to estimate LOS \ion{H}{1} complexity, here we adopt a more restrictive velocity range of $|v|<30$ km/s. If the difference between the depolarization results of the high frequency and low frequency synchrotron datasets is mainly due to significant depth depolarization over the LOS at low frequency, then the smaller velocity-range $N_c^{30}$, which typically probes a smaller portion of the LOS, should affect the high frequency results significantly but not the low frequency one. This is shown in Figure \ref{fig:depol_nc_vrange}, where we compare the depolarization ratio $D$ computed for GMIMS-HBS and GMIMS-HBS binned using $N_c$ with $|v|< 70$ km/s and $|v|< 30$ km/s respectively. The low-frequency results using the two different $N_c$ versions are consistent with each other, while $D$ of the high-frequency dataset drops to a similar level as the low-frequency data when using the smaller velocity-range $N_c$.  Comparing the GMIMS-HBS depolarization ratio distribution binned with $N_c^{30}$ to a null test where we bootstrap resample the $N_c^{70}$ result, we find the difference to be statistically significant with $p<0.001$. One possible interpretation is that the GMIMS-LBS data do not generally probe polarized emission out to the distances of IVCs, as IVCs are included in the $N_c^{70}$ complexity metric, but excluded from $N_c^{30}$.

In general, the depolarization ratio studies presented in this section show a consistent connection between synchrotron depolarization and LOS \ion{H}{1} complexity across multiple frequencies and northern versus southern sky footprints. The variation with frequency can be attributed to increasing depth depolarization at lower frequencies, implying that the high-frequency synchrotron and \ion{H}{1} data probe more similar volumes than the low-frequency synchrotron emission.

\begin{figure*}[t]
    \centering
    \includegraphics[width=0.95\textwidth]{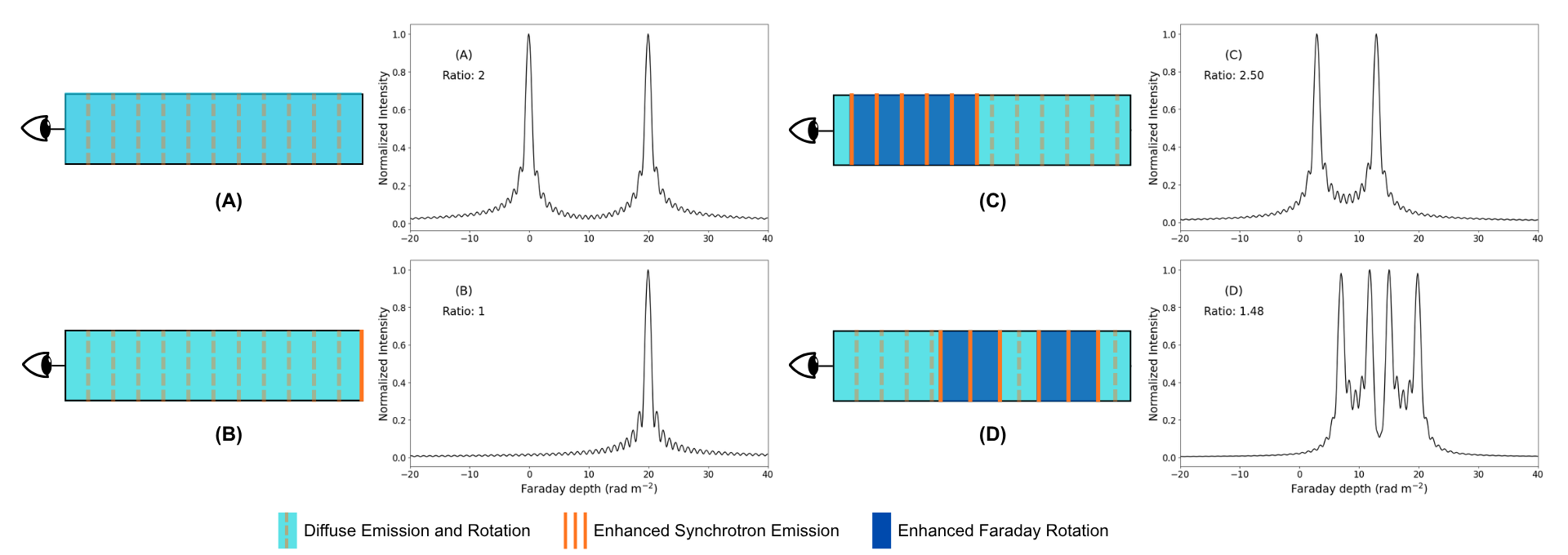}
    \caption{Schematics showing different synchrotron emission+Faraday rotation scenarios and their corresponding Faraday spectra and rotation measure ratio.  (A): Classic Burn slab with uniform diffuse emission and rotation ($R=2$); (B): Rotation entirely foreground to the dominant synchrotron emission contribution ($R=1$). (C) Sightline dominated by a single nearby emitting+rotating dense cloud ($R>2$); (D): Sightline dominated by multiple emitting+rotating dense clouds farther away from observer ($1<R<2$). }
    \label{fig:faraday_spectra_schema}
\end{figure*}

\begin{figure*}[t]
    \centering
    \includegraphics[width=0.95\textwidth]{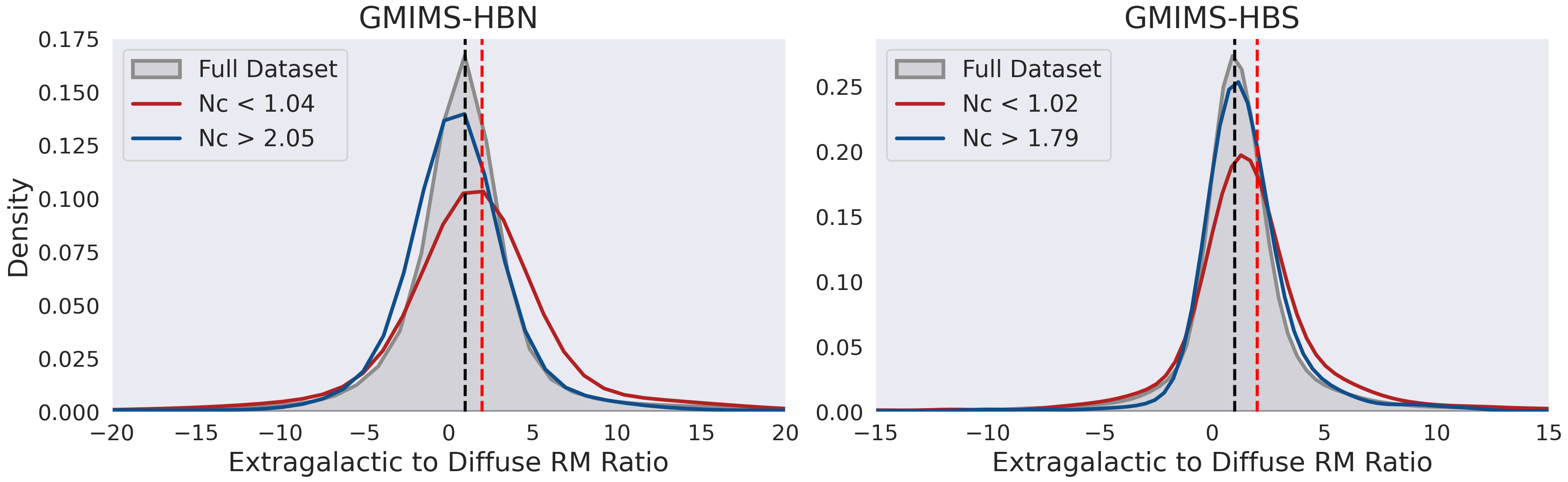}
    \caption{Histogram of extragalactic to diffuse RM ratio in regions of low (10th percentile) vs. high (90th percentile) LOS \ion{H}{1} complexity for GMIMS survey footprints. The black and red dashed line indicates RM ratio of 1 and 2 respectively. Compared to high complexity regions, the RM ratio distribution of the low complexity regions is consistently shifted to larger amplitudes across the northern (left) and southern (right) footprints. }
    \label{fig:rm_ratio_m1}
\end{figure*}

\begin{figure*}[t]
    \centering
    \includegraphics[width=0.95\textwidth]{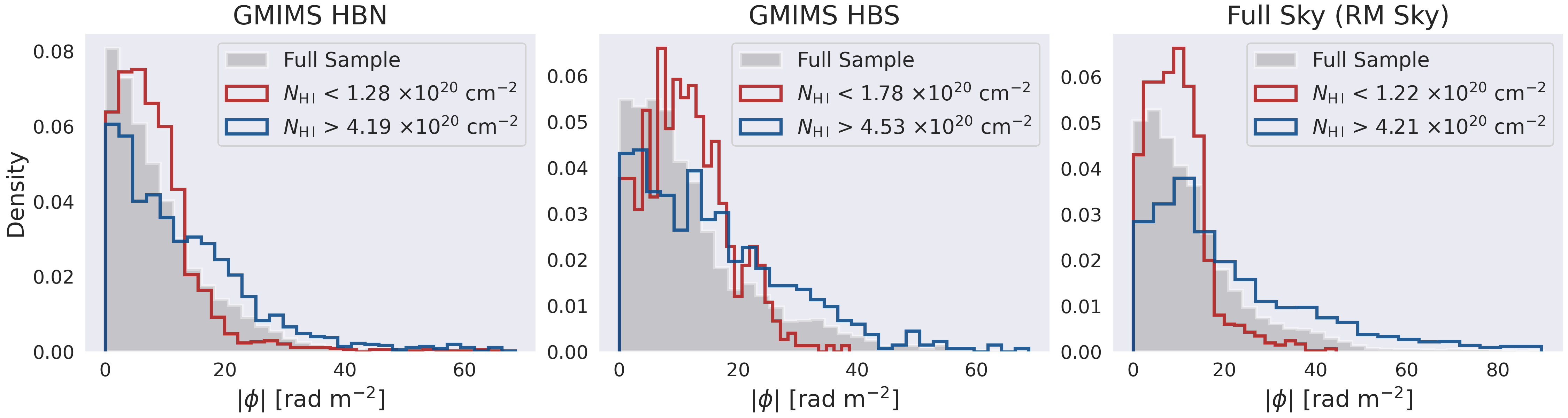}
    \caption{Histogram of the absolute value of the rotation measure distribution in regions of low (10th percentile) vs. high (90th percentile) \ion{H}{1} column density over different dataset footprints. The GHBN (left) and GHBS (middle) first moments results are compared the full sky result (right) using the Galactic rotation measure model from \citet{hutsch22-fm}. Regions of higher \ion{H}{1} column density show consistently higher rotation measure across different footprints.}
    \label{fig:m1_nhi_corr}
\end{figure*}

\section{Effect of \ion{H}{1} complexity On the First Faraday Moment} \label{sec:connect_m1}

The depolarization ratio investigated in Section \ref{sec:fara_hi} used the 0th Faraday moment. However, all three Faraday moments described by Equations \ref{eq:M0}-\ref{eq:M2} contain complementary information about the distribution of the polarized synchrotron emission. Any viable physical explanation for the results in Section \ref{sec:fara_hi} should produce signatures of the synchrotron-HI connection in all three moments in a consistent way. In this section, we examine the first moment $M_1$: the intensity-weighted mean of the Faraday depth $\phi$.

Several recent studies have utilized the ratio of rotation measures between extragalactic sources and extended Galactic emission to probe the large-scale Galactic magnetic field \citep{ordog19-rm, dickey22-rm, erceg22-rm}. For compact extragalactic sources, it is reasonable to assume that the polarized synchrotron emission is dominated by a point source that lies far beyond the Faraday-rotating medium in the Galaxy. In this scenario, the rotation measure can be derived from a simple linear fit to the polarization angle as a function of wavelength: $\psi=\psi_0+\phi\lambda^2$. On the other hand, when measuring extended Galactic emission, we must account for the variation of emitting and rotating medium along the LOS as described by Equation \ref{eq:p_general}. As a result, the ratio between the extragalactic RM and extended emission as measured by the first Faraday moment $M_1$ contains important information about the distribution of diffuse emission and Faraday rotating structures along the LOS. This quantity is known as the ``RM ratio" \citep{ordog19-rm}.

In the simplest case where the emission and rotation are intermixed throughout the LOS with uniform electron density, magnetic field, and synchrotron emissivity, we have the classic ``Burn slab" scenario where the extragalactic to diffuse RM ratio is 2 \citep{Burn66-sb}. Physically, this could correspond to uniform emission and Faraday rotation coming from a pervasive medium like the warm ionized medium (WIM). If instead the Faraday rotating medium is entirely foreground to the bright synchrotron emission, coming from a distant radio source that dominates the diffuse synchrotron emission over the sightline, then the RM ratio is 1. These cases are illustrated more intuitively in panels A and B of Figure \ref{fig:faraday_spectra_schema}. 

Beyond these simple scenarios, fractional values of the RM ratio can arise with more complicated distributions of synchrotron-emission and Faraday-rotating structures along the sightlines. As panel C in Figure \ref{fig:faraday_spectra_schema} illustrates, when the sightline is dominated by a nearby dense cloud with both enhanced synchrotron emission and Faraday rotation (dark blue), relative to the emission and rotation coming from the more pervasive WIM (light blue), the RM ratio can be larger than 2. In contrast, panel D shows a scenario where the RM ratio is between 1 and 2. In this case the sightline is dominated by two smaller emission+rotation enhanced clouds at distances farther from the observer, relative to the WIM background. We hypothesize that these scenarios correspond to our discussion of the connection between synchrotron and LOS HI complexity. Specifically, the \ion{H}{1} cloud structures measured by $N_c$ contribute enhanced synchrotron emission and rotation measure due to enhanced magnetic field strength relative to the WIM. Scenario C corresponds to $N_c\sim1$, where a single nearby \ion{H}{1} cloud dominates the emission+rotation over the LOS. Scenario D corresponds to larger $N_c$ values, where a more distant IVC cloud substantially contributes to the synchrotron emission and Faraday rotation along the sightline. 

We test this hypothesis by examining the extragalactic source to diffuse Galactic RM ratio in regions of different LOS \ion{H}{1} complexity. For extragalactic RM, we utilize the Galactic Faraday rotation sky map from \citet{hutsch22-fm}, which is described in Section \ref{subsec:rm_ex}. The diffuse RM is the first moment $M_1$ derived from the high-frequency datasets GMIMS-HBN and GMIMS-HBS. The much more significant depth depolarization at lower frequencies means that we do not expect the low-frequency GMIMS datasets to correlate well with the extragalactic RM. \citet{dickey22-rm} studied the relation between extragalactic and GMIMS-HBN RM between different latitude ranges, and found strong correlation at $20\degree<|b|<50\degree$ but no correlation at higher latitudes. As a result, we restrict our analysis to this mid-latitude range. 

\begin{figure*}[t]
    \centering
    \includegraphics[width=0.88\textwidth]{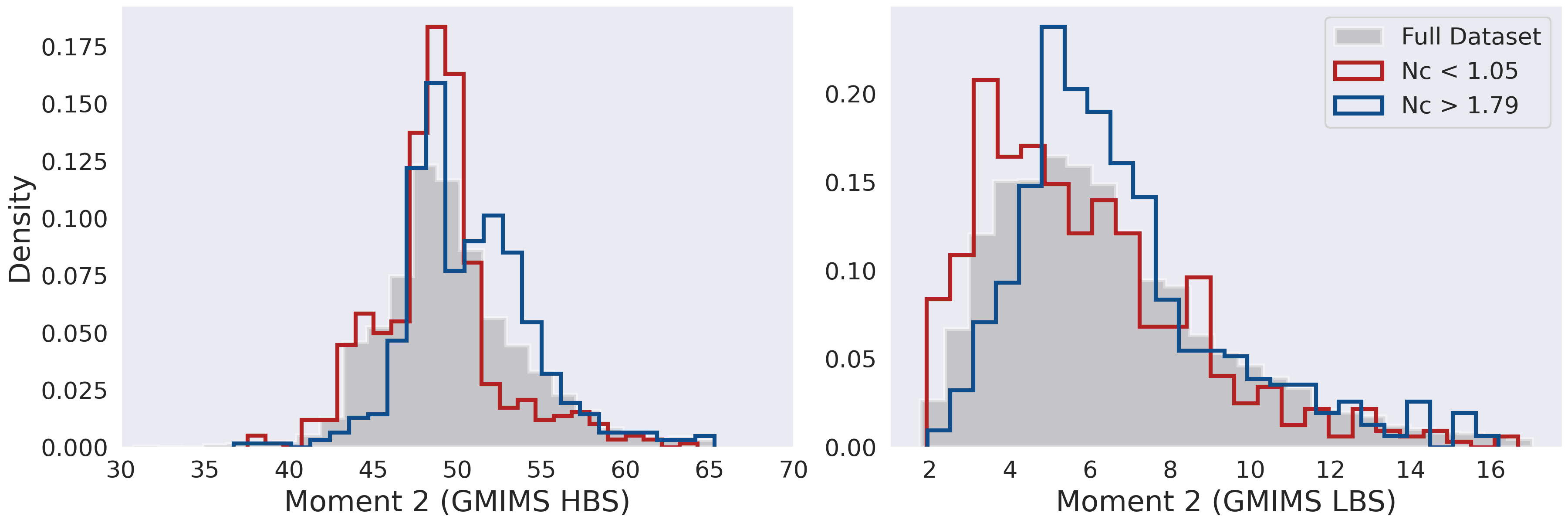}
    \caption{Normalized histogram of the second Faraday moment $M_2$ for the GHBS and LBS datasets in low LOS \ion{H}{1} complexity (10th percentile $N_c$) and high complexity (90th percentile $N_c$) regions. The $M_2$ distribution shifts to higher values in high \ion{H}{1} complexity regions. The effect is more apparent for the low-frequency dataset, which has much better Faraday depth resolution.}
    \label{fig:m2_nc_relation}
\end{figure*}

\begin{figure*}[t]
    \centering
    \includegraphics[width=0.95\textwidth]{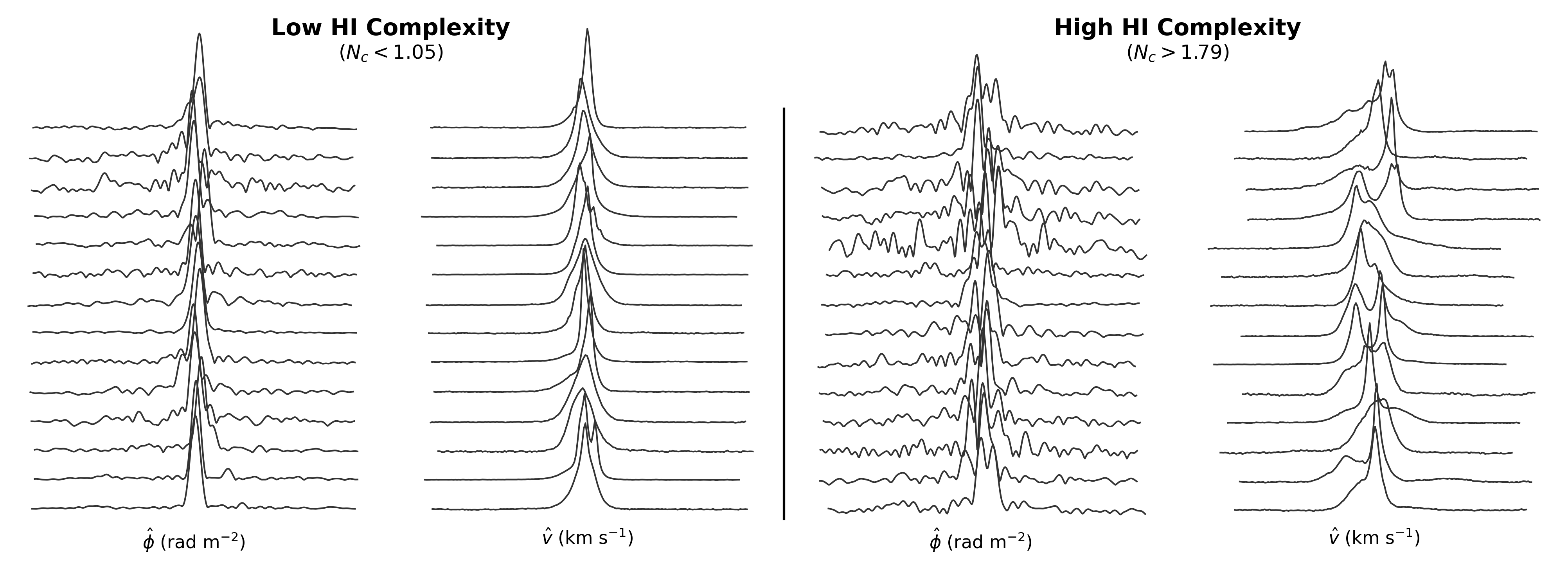}
    \caption{Randomly selected samples of GMIMS-HBS Faraday spectra ($-100 \,\mathrm{rad\,m^{-2}} <\phi<100\,\mathrm{rad\,m^{-2}}$) and \ion{H}{1}4PI \ion{H}{1} spectra ($-120\, \mathrm{km/s}<v<120\,\mathrm{km/s}$) in regions of low (10th percentile, left) versus high (90th percentile, right) LOS \ion{H}{1} complexity. The spectra are normalized to have zero mean and unit standard derivation, and shifted to center at their intensity weighted mean. The Faraday spectra in \ion{H}{1} complex regions generally have more complex spectra distribution.}
    \label{fig:fara_vs_hi_spectra}
\end{figure*}

In Figure \ref{fig:rm_ratio_m1}, we plot the distribution of extragalactic to diffuse RM ratio for GMIMS-HBN, GMIMS-HBS, and combined footprints, in bins of different LOS \ion{H}{1} complexities. The 10th and 90th percentiles are chosen to represent low and high complexity, so that the $N_c$ limits roughly correspond to sightlines dominated by one cloud ($N_c\sim1$) versus sightlines dominated by two clouds ($N_c\sim2$). Compared to the high complexity regions, the RM ratio distribution of the low complexity regions is clearly shifted to larger amplitudes. This is consistent with our hypothesis that \ion{H}{1} complexity traces both synchrotron emitting and Faraday rotating structure complexity, and \ion{H}{1} clouds are likely responsible for both enhanced emission and rotation over diffuse sightlines. We suggest that low complexity ($N_c\sim1$) sightlines correspond to scenarios C in Figure \ref{fig:faraday_spectra_schema} with larger RM ratio, while high complexity ($N_c\sim2$) sightlines correspond to scenario D with smaller RM ratio.

However, there is a caveat to this interpretation. Unlike the simple cartoon scenario in Figure \ref{fig:faraday_spectra_schema}, as the distributions in Figure \ref{fig:rm_ratio_m1} show, the RM ratio value can get arbitrarily large in both positive and negative directions. This is a result of magnetic reversal along the LOS, which can cause extragalactic and diffuse RM to have different signs, or extragalactic RM/diffuse RM sums to $\sim0$ over the LOS \citep{tahani18-rv, dickey22-rm, booth26-rv}. Still, the overall distribution and median value of the RM ratio is consistent with our hypothesis for how \ion{H}{1} complexity traces enhanced synchrotron emission and Faraday rotation.

Another caveat to our interpretation is that the RM ratio result alone is not enough to show that the dense \ion{H}{1} clouds directly contribute enhanced synchrotron emission and Faraday rotation relative to the diffuse emission along the LOS. An alternative interpretation that is still consistent is that the \ion{H}{1} structures simply trace the complexity of the synchrotron emission+Faraday rotation structures, due to some confounding variable that doesn't directly require the enhanced synchrotron emission or enhanced RM to occur within the HI clouds. To more directly examine our first interpretation, in Figure \ref{fig:m1_nhi_corr}, we compare the distribution of $M_1$ in regions of high (90th percentile) and low (10th percentile) \ion{H}{1} column density, using GMIMS-HBN, GMIMS-HBS, and the full-sky Galactic rotation measure model from \citet{hutsch22-fm}. Regions of higher \ion{H}{1} column density show consistently higher rotation measure across different footprints, indicating that more \ion{H}{1} content is directly related to higher rotation measure along the same sightlines. Furthermore, LOS \ion{H}{1} complexity as characterized by $N_c$ is not strongly correlated to total \ion{H}{1} column. The Spearman correlation coefficient between $N_c$ and $N_{H\,I}$ is -0.17 over the diffuse region considered for our study. Therefore, the column density result is not a trivial consequence of the $N_c$ result, and provides additional evidence of a direct association between \ion{H}{1} structure and enhanced Faraday rotation.

\section{Effect of \ion{H}{1} Complexity On the Second Faraday Moment} \label{sec:connect_m2}

The second Faraday moment $M_2$ measures the spread of the Faraday spectrum from its mean value, and encodes information about the complexity of synchrotron-emitting and Faraday rotating structure along the LOS. Generally, higher $M_2$ values do not straightforwardly correspond to the complexity of the Faraday rotating medium along the LOS, and resolving Faraday complexity via $M_2$ is further limited by the Faraday depth resolution of the dataset used \citep{sun15-mc}. As a result, alternative methods of estimating Faraday complexity have been developed, such as QU fitting \citep{miyashita19-qu}, machine learning classification \citep{alger21-ml}, iterative reconstruction \citep{cooray21-ft}, and Bayesian spectral modeling \citep{wenger24-bm}. However, if there is a consistent association between the complexity of synchrotron and Faraday structures and \ion{H}{1} complexity within particular datasets, then we should still expect to observe a connection between higher values of $M_2$ and \ion{H}{1} complexity $N_c$ when averaging over large areas of the sky. 

In Figure \ref{fig:m2_nc_relation}, we show the distribution of $M_2$ for GHBS and LBS datasets in the same low (10th percentile) and high (90th percentile) \ion{H}{1} complexity bins as discussed in the $M_1$ section. In both cases, the $M_2$ distribution shifts to higher values in high \ion{H}{1} complexity regions. This trend is more apparent for the low frequency dataset, since LBS has a much higher Faraday depth resolution at 5.9 $\rm{rad}\ m^{-2}$ compared to HBS at 150 $\rm{rad}\ m^{-2}$.  In Figure \ref{fig:fara_vs_hi_spectra}, we show spectra plots of a random selection of GMIMS-LBS Faraday spectra $P(\phi)$ between -100 $\mathrm{rad\,m^{-2}}$ and 100 $\mathrm{rad\,m^{-2}}$, and HI brightness temperature spectra between -120 km/s and 120 km/s, separated into low (10th percentile) and high (90th percentile) \ion{H}{1} complexity regions. The spectra are normalized and shifted such that they are centered at their intensity-weighted mean value. The normalized Faraday spectra appear visually more simple and aligned in \ion{H}{1} simple regions than in \ion{H}{1} complex regions.

We also examine the $M_2$ distribution of the GHBN dataset, and find that the $M_2-N_c$ relation shows the opposite trend and largely tracks the behavior of $M_0-N_c$, where we found lower $M_0$ values in regions of high $N_c$. This is likely attributed to the lower sensitivity and Faraday depth resolution of the HBN survey. High $M_2$ sightlines in the HBN dataset generally correspond to high-intensity and therefore high $M_0$ sightlines with high SNR. \citet{raycheva25-gm} compared the $M_2$ maps of GHBN and GHBS, and found that GHBN $M_2$ values are generally lower than that of GHBS in overlapping regions, likely attributed to the lower SNR of the GHBN survey. 

\section{Discussion} \label{sec:discussion}


Examining Faraday moment metrics derived from multi-frequency radio polarization surveys, we found a consistent association between Faraday-rotated diffuse synchrotron emission and the measure of LOS complexity probed by \ion{H}{1} emission. First, the relationship between $M_0$ and the LOS \ion{H}{1} complexity, $N_c$, is found to be qualitatively similar to that between dust polarization fraction $p$ and $N_c$, suggesting a common interpretation. In the case of polarized dust emission, the $p$-$N_c$ relationship is naturally explained by depolarization due to different magnetic field orientations in distinct dusty regions along the LOS, where more distinct regions generally means higher $N_c$. As a result, the similar polarized synchrotron emission depolarization trend in high $N_c$ regions  (Figure \ref{fig:p_nc_gmims}) suggests that $N_c$ traces the complexity of polarized synchrotron-emitting structures. 

Secondly, the distribution of the ratio between extragalactic rotation measure and diffuse emission $M_1$ is found to have a higher average in \ion{H}{1} complex ($N_c\sim2$) regions, compared to low-complexity regions dominated by a single \ion{H}{1} cloud ($N_c\sim1$). While the detailed RM ratio distribution is complicated by factors such as magnetic field reversal along the LOS, the association between the overall mean of the distribution and $N_c$ further suggests that \ion{H}{1} complexity probes synchrotron emission complexity, consistent with the $M_0$ result. Specifically, $N_c\sim1$ corresponds to regions dominated by a single, often nearby cloud. If a nearby \ion{H}{1} cloud is also responsible for significant amounts of synchrotron emission, there will be enhanced synchrotron emission closer to the observer relative to the diffuse emission throughout the LOS, resulting in a  higher RM ratio (Figure \ref{fig:faraday_spectra_schema} panel C). In contrast, $N_c\sim2$ sightlines often correspond to sightlines with substantial emission from IVC clouds at much larger distances \citep{panopoulou20-nc}, leading to significant synchrotron emission further away from the observer and a lower RM ratio. Finally, we find an increase in $M_2$ in \ion{H}{1} complex ($N_c\sim2$) compared to \ion{H}{1} simple ($N_c\sim1$) regions. Since $M_2$ is a measure of the variance of the rotation measure spectra, this further suggests that \ion{H}{1} complexity traces the complexity of Faraday rotating features along the LOS. 

\subsection{Interpretation of the Synchrotron-Neutral Complexity Relation} \label{subsec:interpretation}

A natural explanation for the observed synchrotron-neutral connection is that the neutral, partially ionized \ion{H}{1} clouds are directly responsible for significant fractions of the diffuse synchrotron emission and Faraday rotation along the LOS. With respect to polarized synchrotron emission, recent MHD simulation studies \citep{ponnada24-sn, alvarez24-sn, berat26-sm} have argued that the synchrotron emission largely arises from the denser and neutral gas as opposed to the diffuse warm ionized gas, despite the fact that the neutral gas comprises much smaller fractions of the ISM volume. This is explained by the significantly enhanced magnetic field strength in the denser, neutral medium. While other factors such as spatial variation of cosmic ray (CR) density can affect synchrotron emission, by self-consistently evolving cosmic ray energy densities in their simulations instead of assuming simple equipartition with magnetic energy density, \citet{ponnada24-sn} found that while magnetic fields are highly structured and amplified in dense neutral gas, the spatial distribution of CR densities is much more diffuse and smooth. Because diffusive transport and rapid energy losses prevent CR electrons from accumulating in these dense environments, the enhanced synchrotron emission in neutral clouds is driven almost entirely by localized magnetic field amplification. As a result, the dense, neutral medium contributes significant polarized synchrotron emission despite its much smaller volume-filling fraction. Our results can be interpreted as observational evidence for this picture.

For Faraday rotation, recent analyses of LOFAR data showed direct evidence for a connection between LOFAR Faraday structures and the neutral medium, at frequencies $\sim150$ MHz. For example, \citet{boulanger24-wm} estimated the rotation measure associated with the WNM, by using stellar UV observations to derive \ion{H}{1} ionization fraction information. They conclude that WNM electrons and their associated Faraday rotation are foreground to the bulk of the WIM electrons, and make up the main component of the observed Faraday structure at LOFAR frequencies. This is further consistent with results finding morphological associations between \ion{H}{1}/dust and Faraday structures in specific regions of the LOFAR sky \citep{zaroubi15-ni, kalberla17-ni, vaneck17-ph, jelic18-ni, bracco20-ni, ercey24-ft, boulanger24-wm}. At higher frequencies, applying rotation measure synthesis to 563 discrete radio sources between 1.3 GHz and 2.0 GHz observed with the Australia Telescope Compact Array (ATCA), \citet{anderson15-gm} found an association between Faraday complexity and HI column. By explicitly examining LOS HI complexity, our work establishes a more direct connection between Faraday complexity and LOS ISM structures, across a wider range of frequencies and over large areas of the sky. 

The extent to which the Faraday-neutral structure association at higher frequencies can be attributed to the predominantly neutral regions, or explained by confounding factors such as co-spatial ionized and neutral features, and common magnetic field structure, requires extending the detailed physical modeling of \citet{boulanger24-wm} to higher frequencies, which is beyond the scope of the current study. Here we adopt a simple back-of-the-envelope estimation for the feasibility of the partially ionized WNM contributing significant RM to the observed GHBN rotation measure. Over the high latitude diffuse sky, the GHBN has a median $|M_1|\sim7.1 \rm{\,rad\,m^{-2}}$. Making the simplified assumption of a uniformly emitting and rotating slab, this translates to a median $\rm{RM}=2\cdot M_1=14.2\,\rm{rad\,m^{-2}}$. Referencing the $f_{\rm CNM}$ values from \citet{hensley22-cr} and $N_{\rm H\,I}$ from \ion{H}{1}4PI, we estimate an median $N_{\rm WNM}\sim 2.5\times10^{20}\,{\rm cm^{-2}}$ over the same region. Following \citet{boulanger24-wm}, adopting a fiducial \ion{H}{1} ionization fraction estimate of $\sim0.1$ and a helium correction factor of 1.2, the WNM RM equation $\mathrm{RM}_{\mathrm{WNM}}
= 0.26\,\mathrm{rad}\,\mathrm{m}^{-2}
\left(
\frac{N_e^{\mathrm{WNM}}}{10^{18}\,\mathrm{cm}^{-2}}
\right)
\left(
\frac{B_{\mathrm{LoS}}}{\mu\mathrm{G}}
\right)=14.2\,\rm{rad\,m^{-2}}$ implies an electron-density weighted field $|B_{\rm LOS}|$ value of $1.8\,\mu\rm{G}$, within reasonable estimates for typical magnetic field strength in WNM regions \citep{zirnstein16-mf,boulanger24-wm}. While this estimate supports the plausibility of appreciable Faraday rotation within the WNM, the observed association with \ion{H}{1} complexity may also reflect alternative factors such as spatially associated ionized gas or magnetic structure shared by the phases. We next discuss the potential contribution of these factors to the observables presented in the paper.

\subsection{Discussion of Alternative Contributions} \label{subsec:confound_factor}

First, the LOS \ion{H}{1} components characterized by the $N_c$ measure may be spatially and kinematically associated with the WIM structures that are the main source of the observed rotation measure. Such associations can arise from photoionized envelopes surrounding neutral clouds or from ionized and neutral layers within expanding shells \citep[e.g.,][]{haffner01-ia,haffner09-ia,pon14-ia}. In this scenario, $N_c$ indirectly traces the structure of the Faraday-rotating ionized medium, without requiring a substantial Faraday contribution from the predominantly neutral gas itself. To examine this possibility, we utilize the kinematic release from the WHAM Sky Survey, which maps H$\alpha$ across the whole sky between -100 km/s and 100 km/s at 12 km/s resolution. We restrict the data to between -70 km/s and 70 km/s and derive a $N_c^{H\alpha}$ measure using the component counting procedure outlined in \citet{halal24-lb}. We found negligible correlation between $N_c^{H\alpha}$ and \ion{H}{1}-based $N_c$, with a spearman coefficient of 0.06. At the 10th and 90th percentile split of $N_c$, $N_c^{H\alpha}$ values are 1.02 and 1.08 respectively. However, this comparison is limited by the 12 km/s spectral resolution of the WHAM survey. The weak observed correlation therefore does not exclude an association between the neutral structures and the Faraday-rotating ionized medium. Deeper spectroscopy of less thermally broadened lines, such as \ion{N}{2} and \ion{S}{2}, would enable a more discriminating kinematic comparison, while spatially resolved, multi-line surveys such as the SDSS-V Local Volume Mapper can help characterize the interfaces between the phases \citep{drory24-ph}. 

Secondly, the observed synchrotron and LOS \ion{H}{1} complexity association could be attributed to common large-scale magnetic field structures. In particular, the mean orientation of the Galactic Magnetic Field (GMF) in the solar neighborhood with respect to the LOS can imprint large-scale patterns on the observed polarized sky at synchrotron- and dust-dominated frequencies \citep{mdma08-gm, planck15-me}, affecting the observed polarization statistics. However, over the diffuse high latitude sky examined in this paper, we do not expect a direct relationship between local GMF structures and the LOS complexity probed by $N_c$. \citet{aghanim16-pk} examined polarization statistics of Planck 353 GHz emission over the Southern Galactic cap region ($b<-60\degree$), and found that a GMF model with an ordered field and a turbulent component fails to reproduce the observed polarization fraction distribution, particularly its low-polarization peak and high-polarization tail. Instead, including field-orientation along the LOS, represented by multiple independent polarization layers, allows the model to reproduce both polarization angle dispersion and the polarization fraction distribution. \citet{adak20-md} examined the Northern Galactic gap with a similar magnetic field model, incorporating three \ion{H}{1} column density templates as a proxy for varying dust emission layers. These studies indicate that for high latitude regions, a large-scale GMF model with ordered + turbulent components alone cannot account for the statistics of the observed dust polarization fraction distribution without accounting for depolarization due to multiple LOS components. 

To directly test the effect of a large-scale ordered magnetic field, we derive a geometric depolarization factor $g$ due to large-scale GMF orientation with respect to the LOS, following the ordered+turbulent component field model outlined in \citet{mdma08-gm}. We do not find a strong relationship between $g$ and $N_c$, with a spearman coefficient of -0.19 over the high latitude sky. Furthermore, the depolarization ratio result presented in Section \ref{subsec:fara_hi_freq} remains consistent within 10\% when adding $g$ as an additional confounding variable in addition to total intensity $I$. As a result, large-scale patterns due to the orientation of the local ordered field with respect to the LOS alone cannot account for the observed relationship with $N_c$. However, this does not rule out connected magnetic field structure being a contributing factor in the observed synchrotron and $N_c$ association. Specifically, large-scale patterns are present in the $N_c$ map because higher $N_c$ is associated with the presence of IVC gas \citep{panopoulou20-nc}. As many IVCs reside at the disk-halo interface \citep{richter17-iv, lehner22-iv}, the association between higher $N_c$ and a greater IVC contribution could reflect sightlines sampling a transition in magnetic-field structure between the disk and halo. However, a generic disk–halo magnetic field structure transition would not by itself explain the observed association with $N_c$, since both low and high complexity sightlines traverse this transition. An explanation without assuming substantial Faraday rotation within the predominantly neutral gas itself would additionally require the IVC locations to directly correlate with changes in magnetic field geometry, or enhanced RM contribution from ionized structures associated with IVCs. Utilizing WHAM data, \citet{stil16-iv} find that H$\alpha$ emission from an intermediate-velocity filament is associated with excess RM of background extragalactic sources. However, this result concerns a single ionized filament whose association with a neutral IVC complex remains uncertain. It would be valuable for follow-up work to apply our $N_c$ and Faraday-moment analysis to synthetic \ion{H}{1} and polarized synchrotron observations from multiphase MHD simulations \citep[e.g.,][]{ponnada24-sn, alvarez24-sn}. Tracking the synchrotron emissivity and Faraday rotation contributed by each gas phase, together with controlled tests of magnetic-field geometry, would help distinguish contributions from predominantly neutral gas from those of ionized structures spatially and kinematically associated with IVCs, while assessing the role of magnetic structure shared by the phases at the disk--halo interface. 


\section{Conclusion} \label{sec:conclusion}

In this work, we present a comprehensive analysis of the relationship between the complexity of the neutral interstellar medium and the properties of polarized synchrotron emission and Faraday rotation.  By comparing \ion{H}{1} LOS complexity ($N_c$) with Faraday moments derived from GMIMS and other surveys, we have demonstrated a consistent association that persists across multiple frequency datasets ($\sim$300 MHz -- 23 GHz) over the high latitude sky. Our primary findings are summarized as follows: 
\begin{itemize}
    \item We devise a series of tests that compare the statistical behavior of radio polarization data toward sightlines dominated by a single HI cloud (``simple" sightlines) to those with multiple \ion{H}{1} clouds with similar column densities (``complex" sightlines). The simple sightlines tend to probe nearby \ion{H}{1}, while complex sightlines tend to intersect IVC structures \citep{panopoulou20-nc}.
    \item Similar to how dust polarization at 353 GHz decreases in regions of increasing LOS \ion{H}{1} complexity, we observe depolarization in \ion{H}{1}-complex regions across frequencies dominated by polarized synchrotron emission. This suggests that \ion{H}{1} complexity traces the complexity of diffuse polarized synchrotron emission structures. 
    \item We observe higher degree of depolarization in \ion{H}{1} complex regions for high frequency ($>$ 1.4 GHz) synchrotron emission datasets compared to low frequency ($\sim$ 500 MHz) ones, which we attribute to greater degree of depth depolarization at low frequencies. 
    \item By restricting \ion{H}{1} velocity range from $|v|<70$ km/s to $|v|<30$ km/s, we derive an alternative LOS complexity measure $N_c^{30}$ that probes a smaller portion of the sightline. We found that the depolarization ratio computed with $N_c^{30}$ decreases significantly for high frequency but not for the low frequency synchrotron datasets, supporting the hypothesis that the high frequency synchrotron emission probes the complexity of a larger portion of the sightline out to greater distances than the low frequency emission due to frequency-dependent depth depolarization. 
    \item Utilizing ratios of extragalactic RM to diffuse emission RM to probe the LOS distribution of synchrotron emitting and Faraday rotating structures, we find that the RM ratio increases in \ion{H}{1} simple regions dominated by a single nearby cloud, compared to \ion{H}{1} complex regions dominated by distant IVC clouds. This further supports an association between the complexity of \ion{H}{1} and synchrotron emission structures. 
    \item Probing Faraday complexity using the second moment $M_2$ of the rotation measure spectra, we find that $M_2$ increases in \ion{H}{1} complex regions, suggesting a direct connection between Faraday complexity and \ion{H}{1} complexity.
\end{itemize}
The different Faraday moment analyses, each probing a different aspects of the Faraday spectra, show consistent association between Faraday-rotated diffuse synchrotron emission and LOS \ion{H}{1} complexity measures. With the advancement of 3D dust mapping techniques \citep{zhang23-dm,edenhofer24-dm}, and upcoming starlight polarization surveys \citep{tassis18-ps, pelgrims24-pa, tahani26-df}, our results provide a blueprint to synthesize 3D dust/\ion{H}{1} information with polarized synchrotron surveys to probe the 3D multiphase ISM and magnetic field structure.

\bigskip
\noindent We thank Marc-Antoine Miville-Desch{\^e}nes, Marijke Haverkorn, Alex Hill and Anna Ordog for helpful discussions. This work was supported by the U.S. National Science Foundation under CAREER grant No. AST-2441452, and S.E.C. acknowledges additional support from an Alfred P. Sloan Research Fellowship. M.L. acknowledges support from a 2MB Graduate Fellowship for Frontier Research. A.S. is supported by the Australian Research Council through the Discovery Early Career Researcher Award (DECRA) Fellowship (project~DE250100003) funded by the Australian Government and the Australia-Germany Joint Research Cooperation Scheme of Universities Australia (UA--DAAD, 2025--2026).

This work benefited from the conference ``Structure and polarization in the interstellar medium: A Conference in Honor of Prof. John Dickey", a hybrid meeting hosted jointly at Stanford University and at the Australia Telescope National Facility in February 2025. We acknowledge support from the National Science Foundation (NSF Award No. 2502957), from the Kavli Institute for Particle Astrophysics and Cosmology, from the Commonwealth Scientific and Industrial Research Organisation, and from the Australian Research Council. The authors acknowledge Interstellar Institute’s program “II6” and the Paris-Saclay University’s Institut Pascal for hosting discussions that nourished the development of the ideas behind this work.

%



\software{astropy \citep{astropy13}, 
          NumPy \citep{numpy11},
          scipy \citep{scipy20},
          healpy \citep{healpy05-hp},
    }





\bibliography{farahi_complex}
{}
\bibliographystyle{aasjournal}



\end{document}